\documentclass[a4paper,11pt,twoside]{llncs}
\usepackage{graphicx}				%F�r \includegraphics
\usepackage{url}				%Um URLs so anzugeben, dass daraus Links werden
\usepackage[breaklinks=true,pdfhighlight=/P,pdfmenubar=true,pdftoolbar=true,pdfpagelabels=true,pdfstartpage=1,pdfstartview=FitV,pdftitle={An.Example.of.Complex.Pedestrian.Route.Choice},pdfsubject={An.Example.of.Complex.Pedestrian.Route.Choice},pdfauthor={Kretz},pdfcreator={Kretz},pdfproducer={Kretz},pdfkeywords={pedestrians,routechoice}]{hyperref}	%Macht aus Referenzen interne und externe Links
\usepackage[numbers,sort&compress]{natbib}	%Sortiert Zitate und macht aus [10,11,12,13,14] [10-14]
\usepackage[american]{babel}		%Sprachbibliotheken: Neue deutsche Rechschreibung und amerikanisches Englisch (Trennregeln!)
\begin{document}
%\selectlanguage{american}

%
%HAUPTTEIL%%%%%%%%%%%%%%%%%%%%%%%%%%%%%%%%%%%%%%%%%%%%%%%%%%%%%%%%%%%%%
\title{An Example of Complex Pedestrian Route Choice}
%\toctitle{An Example of Complex Pedestrian Route Choice}
\author{Florian Gr{\"a}{\ss}le and Tobias Kretz}
\institute{PTV AG, Stumpfstra{\ss}e 1, D-76131 Karlsruhe\\\email{\{Tobias.Kretz\}@ptv.de}}
%\author{Florian Gr{\"a}{\ss}le and Tobias Kretz\\ \\PTV Planung Transport Verkehr AG \\ Stumpfstra{\ss}e 1 \\ D-76131 Karlsruhe\\\tt{\{Tobias.Kretz\}@ptv.de}}

\maketitle

\begin{abstract}
Pedestrian route choice is a complex, situation- and population-dependent issue. In this contribution an example is presented, where pedestrians can choose among two seemingly similar alternatives. The choice ratio is not even close to being balanced, but almost all pedestrians choose the same alternative. A number of possible causes for this are given.

\end{abstract}

\section{Introduction}
Route choice of pedestrians in general is complex for a number of reasons. Pedestrians can stick to a more or less clearly defined graph of links and nodes given by city planners or they can ignore these bituminized suggestions for route choice and plan their path in 2d, passing any kind of surface, and deviating from air line only when they are forced by insuperable obstacles. In reality the walking behavior will be determined by something inbetween, as links never are strictly one-dimensional objects and -- even more important -- there are lots of obstacles and conditions, which are not insuperable, but just a bit unpleasant to pass over or by.

A second issue, which is more complicated for pedestrians than for vehicular traffic, is the set of determinant factors for route choice. Most vehicle drivers strongly prefer the quickest route. For pedestrians the quickest route bears some attractiveness as well, but sometimes they may find a more scenic or a safer route or a route with paved tracks more attractive and decide to go for it. It's an almost infinite task to figure out, for which groups of the population under which circumstances a given ``amount of safety'' equals what increase of travel time. While the task appears difficult, it is not impossible. 

In combination these two issues are that complex that they immediately raise the question on the impact of misperceptions and false estimations. How many pedestrians name a route the shorter one, if it is by 1\% longer than the truly shortest route? How many pedestrians are able to find the correct safest route, compared to accident report data? And how many traffic planners have comparable scenic and esthetic preferences as the general population has?

After all there is one simplification for pedestrians as participants of traffic compared to vehicle drivers: the quickest path most of the time only slightly deviates from the shortest, as pedestrian densities in public space normally are fairly small. The situation changes for events, where large crowds gather intentionally. There the quickest path can deviate considerably from the shortest \cite{Kretz2009,Meschini2009}.

This work reports about an observation, where pedestrians in large numbers in public space refrain from choosing the shortest (and quickest) route. There are only two main alternative routes and the geometry is -- at first sight -- almost symmetric. The deviations from perfect symmetry appear to be rather small, still they seem to have a strong impact on route choice behavior and spawn a large set of causes for the deviations from the choice of the shortest path.

The outline of the paper is as follows: first the geometry and situation are described, then the observations are reported and finally causes for the observed behavior are discussed.

\section{Motivation}
Probably the first thing that comes to mind, why understanding pedestrian route choice would be necessary, is the planning of large crowd events in geometrically non-trivial environments. But understanding what is considered to be element of an attractive route also helps to build cities that are considered to be attractive by the local population and thus create activity and even have a positive effect on the economic development of a city \cite{Chang2002,Sarkar2003}. And knowing which elements are helpful for orientation and navigation enables city planners to assist tourists -- and strangers in general -- to find their own way through the city \cite{_Lynch1960,OConnor2005}.

If it is justified to assume that the observed pedestrians know about the attractiveness and utility of the available route alternatives, the observed route choices give the actually preferred routes and thus the resulting data can be used to calculated route suggestions that meet the expectations of a user of a pedestrian navigation device \cite{Millonig2006,Millonig2008,Millonig2010}.

To assess the value of a shop location in new retail trade infrastructure it is crucial to know about preferred routes to estimate the number of walk-in customers \cite{Borgers1986,Haupt2006}.

\section{Geometry and Situation}
Karlsruhe's main soccer stadium -- the ``Wildparkstadion'' -- until today is neither connected directly to Karlsruhe's tram system, nor does it have sufficient parking spaces, which makes the vehicular traffic situation before and after matches of the city's largest soccer club KSC quite inconvenient. So, many fans decide to approach by tram and walk the remaining about 2 km to the stadium. For this they are using a number of stations to leave the tram. Why someone chooses a specific station would be worth a different study, here we focus on those fans, who walk from station ``market place'' to the stadium.

\begin{figure}[htbp]
  \center
	\includegraphics[width=0.618\textwidth]{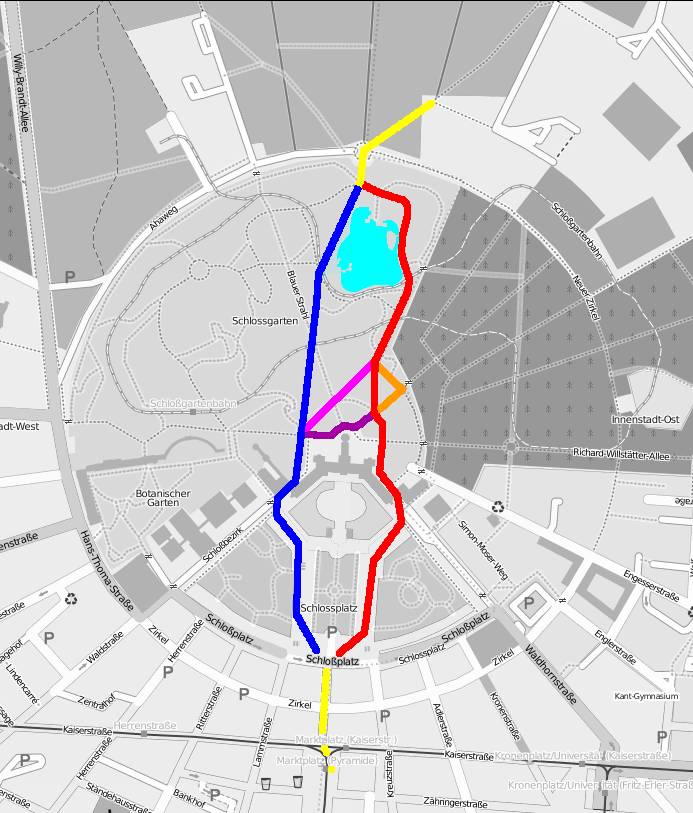}
	\caption{The basic graph of routes emerging from the fans' movements.\cite{Openstreetmap2009}}
	\label{fig:openstreetmap}
\end{figure}

Karlsruhe's market place is located almost exactly south of the central tower of Karlsruhe castle, which itself is built symmetrically approximately 135 m each to east and west referred to the axis castle tower $\leftrightarrow$ market place. Therefore the castle poses an obstacle for pedestrians, who want to walk northbound from market place to -- for example -- the stadium. Its function as an obstacle is weakened by one gate each in the castle's east as well as west wing about 77 m from the symmetry axis. These two gates are ``fix points'' for the soccer fans: only a few meters north of market place each fan has to decide, if he wants to pass the east or the west gate.

Between the market place and the castle (i.e. south of the castle) lies the castle square. In the castle square there is no notable deviation from a perfect symmetry related to named symmetry axis. This implies that the walking distance from market place to each of the two gates for practical purposes is identical.

North of the castle the castle garden extends. Originally it was also laid out in an almost perfect symmetry in 1721, but this was given up around 1800. There is now a pond, which stretches a bit more to the east than the west (the cyan spot in figure \ref{fig:openstreetmap}). For this reason passing the pond on the west is the shorter and quicker route. East of the pond there is a paved walkway that leads northwards in a somewhat zig-zag-style. West of the pond there are no walkways, at least not next to the pond, just a well trimmed meadow.

At the north tip of the nearly circle shaped area of castle park and castle garden there is one single gate (the ``north gate'') exactly on the symmetry axis, which all fans have to pass. After this they have to turn north east to walk straightly to the stadium (the upper yellow route in figure \ref{fig:openstreetmap}).

From the point, where the fans have to decide, if they want to walk the eastern or the western route to the north gate, 895 m have been measured for the western route (the blue route in figure \ref{fig:openstreetmap}) and 926 m for the eastern route (the red route in figure \ref{fig:openstreetmap}). The measurement has been done using a GPS device and simply walking both routes, trying to minimize the distance along each route wherever possible. The difference of 31 m is seemingly small. However, it is possible to correctly estimate the shorter route without tools -- as a matter of fact realizing this difference was the initiation for this study and for measuring the distance with a tool. 

\section{Observations}
\begin{figure}[htbp]
  \center
	\includegraphics[width=0.618\textwidth]{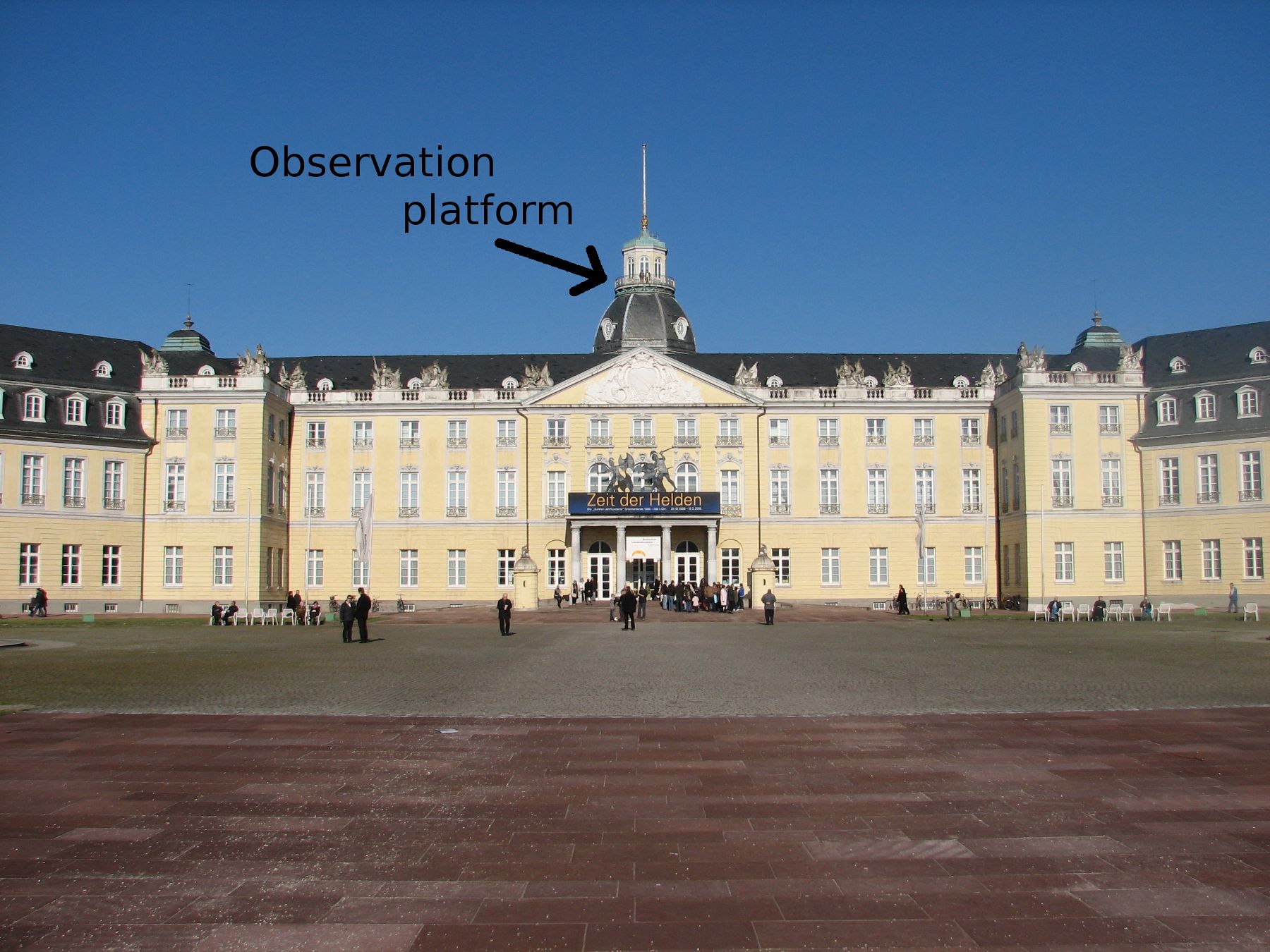}
	\caption{The castle tower served as observation platform.}
	\label{fig:castle}
\end{figure}

The castle tower is accessible for the public and was ideal to observe the flows of fans (see figure \ref{fig:castle}).
\begin{figure}[htbp]
  \center
	\includegraphics[width=0.618\textwidth]{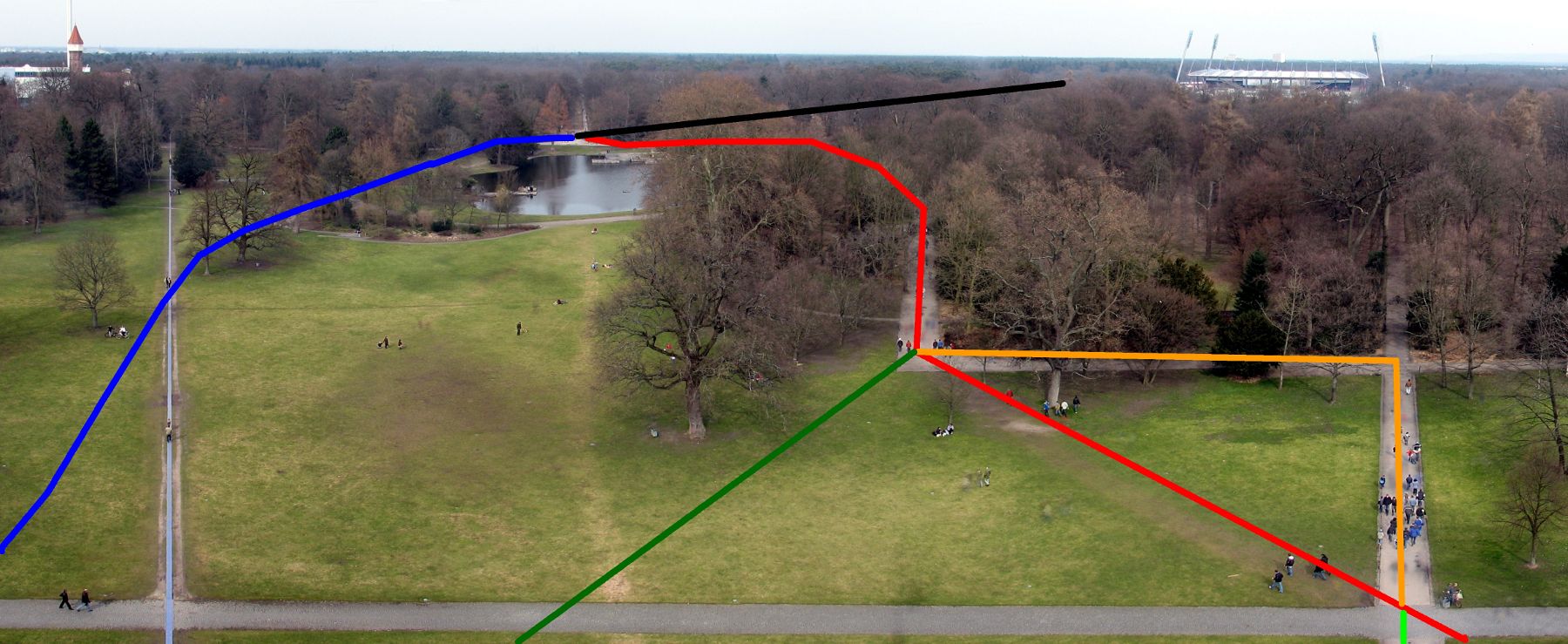}
	\caption{Viewing north from the castle tower. The colors of routes correspond to those of figure \ref{fig:openstreetmap}. Note that this is a panoramic image and therefore the radial paths almost appear as parallels and the circle as a horizontal path.}
	\label{fig:north}
\end{figure}
\begin{figure}[htbp]
  \center
	\includegraphics[width=0.618\textwidth]{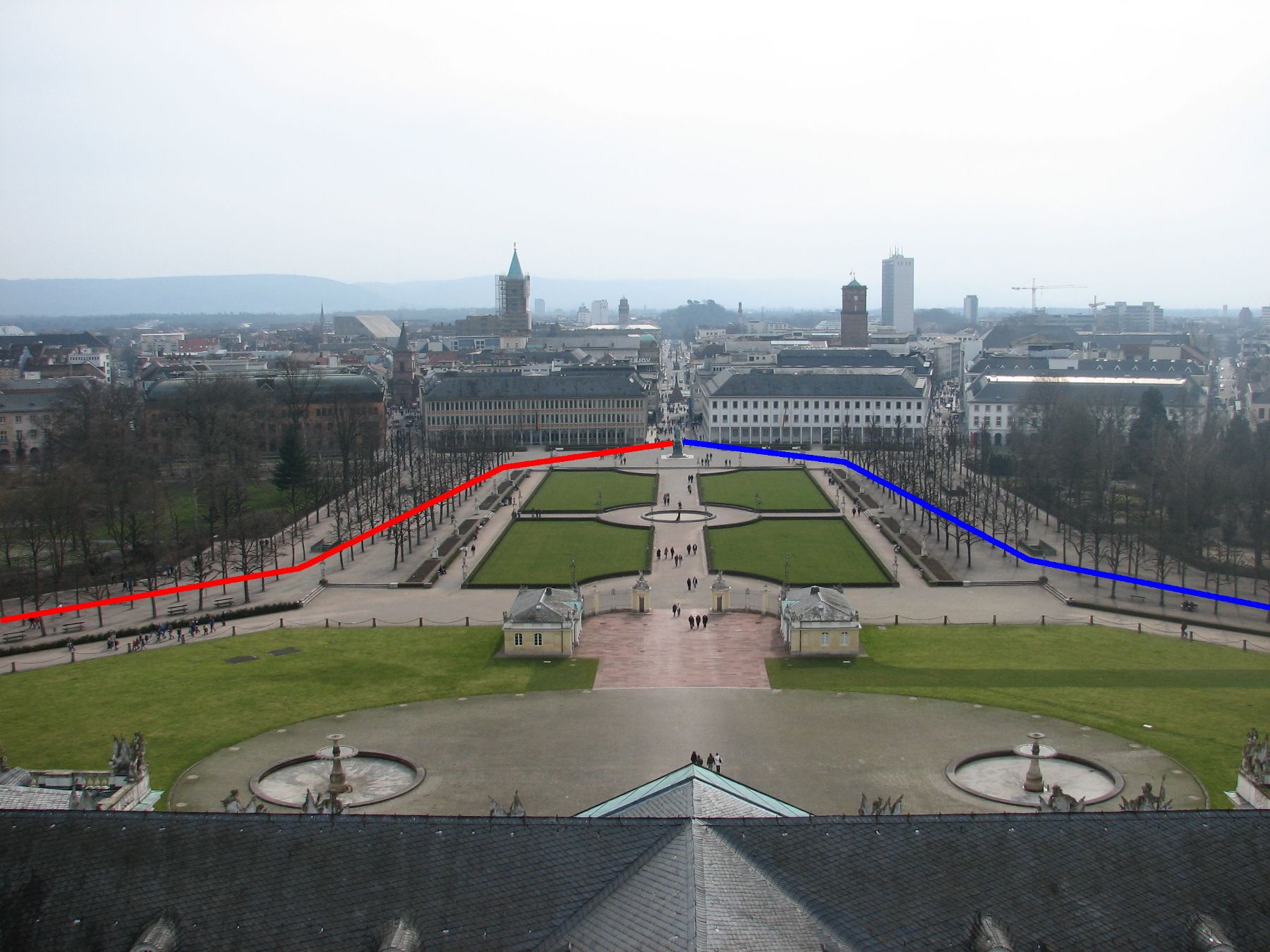}
	\caption{Viewing south from the castle tower. The colors of routes correspond to those of figure \ref{fig:openstreetmap}. }
	\label{fig:south}
\end{figure}
Almost 100\% of the fans pass the pond on the east, i.e. on the longer route. Similarly almost 100\% entirely (also south of the castle) walk the eastern route. There are even fans entering the castle garden through the western gate (they do not necessarily come from market place, but also from western city quarters) and cross the symmetry axis south of the pond to continue their way on the eastern route. Some fans on the eastern route accept an additional detour to walk the whole route on a paved track.

After the match, when the fans return from the stadium to the market place a large share of fans passes the pond on the western side. But nearly 100\% of them immediately after the western gate turn west and do not walk towards the market place, but pass the Supreme Court building and head for Western Town. The tiny volume of fans walking from western gate to market place can be explained by people, who have walked with friends from the north gate to western gate and then parted way, as their friends headed for Western Town.

\section{Discussion}
When asking, why almost no fan is choosing the shortest (and quickest) route, one has to bear in mind that a large share of fans walks this way frequently and knows the area quite well. On their way to the stadium time does not matter a lot and also not an extra 30 meters of distance. This may change after the match, when people are tired and maybe in a hurry to reach for home or a bar. Still, of those fans, who reach for market place, almost noone walks the shorter western route. This is even more astonishing as those fans, who pass the pond on the western side (as in the end they reach for Western Town) show that there is a viable track. In addition on the way back -- contrary to walking toward the stadium -- the fans face the castle tower, which is a good landmark that helps to walk the shortest track over the grass to the western gate (compare \cite{_Lynch1960}). Thus a vast majority, if not all, fans who head for market place either erroneously estimate the eastern route to be the shorter one, or -- despite of having reasons to walk the western route after the match than before -- still see utilities in the eastern route that more than balance the 30 meters that the western route is shorter.

From just observing the fans one cannot tell the true reason. It's even doubtful that a reliable answer could be found with questionnaires. Such decisions are often done more or less unconsciously and when being asked, people often wonder about the reasons themselves. Therefore here possible reasons are listed, which in their diversity demonstrate the complexity of pedestrian route choice, even in this seemingly manageable almost ``laboratory'' example.

\subsection{False Estimation}
In the castle garden the eastern route is paved and the western route not. Moreover the eastern route is longer for its zig-zag character. For these reasons walking on the western route demands more intellectual activity and the eastern route appears to be better structured and as a sequence of short pieces. When remembering the walked route these two facts actually can make the eastern route appear shorter than the western route.

\subsection{Utility and History}
On their way to the stadium one can observe fans on the eastern route, who accept an additional detour to avoid walking over grass at all (the orange route in figure \ref{fig:openstreetmap}). And this on a very clear area, where both alternatives are entirely visible. This was observed on a dry day, when there was no danger of becoming dirty walking off the paved tracks. Obviously for this share of fans it's clear that they value a paved track a lot and accept large detours, if they can avoid leaving it. For other fans this might not be that an important factor, but still be significant, if it's about larger distances. A comparable phenomenon are fans entering the castle garden by the western gate to cross the symmetry axis (the light or dark magenta route in figure \ref{fig:openstreetmap}) to proceed on the eastern route.

A strong influence on route choice of commuting pedestrians -- as one can consider many of the fans to be -- is to walk the same route as last time. For this reason one has to consider the utility of a paved track compared to an unpaved one on a rainy day and in the dark, even though the observations were made on dry and bright days. Everyone easily can feel the disutility that a route has that leads over a meadow with extended poddles. During winter time, when it's already dark, when the fans walk back home from the stadium, the bright concrete of the paved eastern route gives guidance in addition to the safety that one will most probably not sprain one's foot, as one might do in a hole in the meadow on the western route. And finally even in summer there is an additional disutility for walking the western route: the meadow one has to walk over on the western route is filled with people relaxing in the sun. One has to do a number of small detours and probably will lose the time advantage one in principle could gain.

\subsection{General Direction}
When the fans on the way to the stadium have to choose between eastern or western route, the stadium lies to the east. If -- without thinking much about route choice consciously -- they ignore the fact that they have to pass the northern gate, it might seem the natural choice to choose the direction, which is most similar to the direction of linear distance. This argument wouldn't be valid for the return route. For this one could only assume that, if in doubt, most people choose the same route as when approaching the stadium.

\subsection{Social Attachment instead of Equilibrium}
It can be assumed that for the fan flows there is a tendency to strengthen majority choice and weaken minority choice, i.e. whatever ratio between western route and eastern route there once was, it naturally drove toward 0\%:100\%. Not only groups stick together, where the individuals know each other, but especially the way to the stadium is ``celebrated''. Even strangers seek the proximity of others, if they notice these others as being fans as well. This attraction does not only drive route choice ratios to the extreme, but can also initially be the cause for the observed phenomenons: there is a further tram station east of the market place and there is a famous fan bar in the Eastern Inner City. For fans coming from this second station or walking back to this station and fans walking to that bar after the match in the castle garden the eastern route is the shortest and quickest one. These fans pose an attraction for those coming from or going to market place. However, as already mentioned a similar effect exists with fans walking to the western city quarters, but the volume of fans there might be smaller.

This phenomenon of social attraction is opposite to the effect that drives a traffic system toward user equilibrium. As the travel time on a given route grows monotonously with density, the individual driver has to balance the factors short route vs. low density, i.e. it's a repulsive ``force'' that's responsible for the dynamic user equilibrium. 

The social attachment model \cite{Mawson1980,Mawson2005} claims that people in danger seek to come close their friends or just other people. In an extreme form this behavior is often termed ``herding behavior''. Thus, while in such situations travel time matters, these effects for one instance may make people choose not the shortest route and additionally may produce an unbalanced route load, leading to further delays -- especially leading to anything but a state close to user equilibrium.

For the soccer fans on their way to the stadium social attachments might simply be more important than a few saved seconds. The density on the route is that low anyway, that everyone can maintain the desired speed. On the return path there are already reasons stated above, why travel time may matter more. Additionally the density is higher and on the way to the northern gate not everyone can walk freely. This might even still be true for those walking the eastern route. As the western route has no definite width, overtaking is easy and the difference in travel times between eastern and western route becomes larger. Still, as stated above almost no fans heading for market place do choose the western route.

\section{Summary}
In this contribution the pedestrian flow ratios in a scenario with a rather simple geometry were investigated. The observations were astonishing at first, if expectations are based on typical experiences with vehicular traffic. There is a number of reasons to explain the observations. But none could be identified definitely to be the decisive one; and the last issue (no fans on the western route heading for market place) remains more or less open, if one does not want to merely invoke ``habits''. These open questions together with the special geometry of the location suggest further investigation of the situation, including for example questionnaires, more quantitative evaluations, and observations under more diverse circumstances (weather and lighting conditions).

%
%BIBLIOGRAPHY%%%%%%%%%%%%%%%%%%%%%%%%%%%%%%%%%%%%%%%%%%%%%%%%%%%%%%%%%%
\nocite{_PED2008}
\bibliographystyle{utphys_quotecomma}
\bibliography{KSC}

\providecommand{\href}[2]{#2}\begingroup\raggedright\begin{thebibliography}{10}

\bibitem{Kretz2009}
T.~Kretz, ``{Pedestrian Traffic: on the Quickest Path}'',
  \href{http://dx.doi.org/10.1088/1742-5468/2009/03/P03012}{{\em Journal of
  Statistical Mechanics: Theory and Experiment} {\bf P03012} (2009)  },
  \href{http://arxiv.org/abs/0901.0170}{{\tt arXiv:0901.0170
  [physics.soc-ph]}}.

\bibitem{Meschini2009}
L.~Meschini and G.~Gentile, ``{Simulating car-pedestrian interactions during
  mass events with DTA models: the case of Vancouver Winter Olympic Games}'',
  in {\em SIDT 2009 International Conference - The effects of important events
  on land-use and transport: towards Milan Expo 2015 and Naples Forum 2013,
  Milan, Italy}.
\newblock 2009.

\bibitem{Chang2002}
D.~Chang, ``{Spatial choice and preference in multilevel movement networks}'',
  \href{http://dx.doi.org/10.1177/0013916502034005002}{{\em Environment and
  Behavior} {\bf 34} (2002) no.~5, 582--615}.

\bibitem{Sarkar2003}
S.~Sarkar, ``{Qualitative evaluation of comfort needs in urban walkways in
  major activity centers}'', in {\em {TRB annual meeting}}.
\newblock 2003.

\bibitem{_Lynch1960}
K.~Lynch, {\em {The image of the city}}.
\newblock MIT press, 1960.

\bibitem{OConnor2005}
A.~O'Connor, A.~Zerger, and B.~Itami, ``{Geo-temporal tracking and analysis of
  tourist movement}'',
  \href{http://dx.doi.org/10.1016/j.matcom.2005.02.036}{{\em Mathematics and
  Computers in Simulation} {\bf 69} (2005) no.~1-2, 135--150}.

\bibitem{Millonig2006}
A.~Millonig and K.~Schechtner,
  \href{http://dx.doi.org/10.1007/978-3-540-47064-9_10}{``{Decision Loads and
  Route Qualities for Pedestrians-Key Requirements for the Design of Pedestrian
  Navigation Services}'',} in {\em Pedestrian and Evacuation Dynamics 2005},
  N.~Waldau, P.~Gattermann, H.~Knoflacher, and M.~Schreckenberg, eds.,
  pp.~109--118.
\newblock Springer Verlag, 2006.
\newblock ISBN: 978-3-540-47062-5.

\bibitem{Millonig2008}
A.~Millonig and G.~Gartner, ``{Exploring Human Spatio-Temporal Behaviour
  Patterns}'', in {\em AutoCarto 2008, The 17th International Research
  Symposium on Computer-based Cartography Shepherdstown, West Virginia, USA}.
\newblock Cartography \& Geographic Information Society, 2008.

\bibitem{Millonig2010}
A.~Millonig and G.~Gartner,
  \href{http://dx.doi.org/10.1007/978-3-642-04504-2_50}{``{A Multi-Method
  Approach to the Interpretation of Pedestrian Spatio-Temporal Behaviour}'',}
  in Klingsch {\em et.~al.} \cite{_PED2008}.
\newblock ISBN: 978-3-642-04504-2.

\bibitem{Borgers1986}
A.~Borgers and H.~Timmermans, ``{A Model of Pedestrian Route Choice and, Demand
  for Retail Facilities within Inner-City Shopping Areas}'', {\em Geographical
  analysis} {\bf 18} (1986) no.~2, .

\bibitem{Haupt2006}
T.~Haupt, ``{Verkehrs- und Raumplanung –- Wirkungssimulation –- Information:
  Praxis und Vision}'', in {\em 11th Conference on Urban Planning and Regional
  Development in the Information Society}, M.~Schrenk, ed., pp.~543--554.
\newblock Vienna, 2006.
\newblock
  \url{http://www.corp.at/corp_relaunch/papers_txt_suche/CORP2006_HAUPT.pdf}.
\newblock (in German).

\bibitem{Openstreetmap2009}
Openstreetmap, ``{Openstreetmap -- Karlsruhe Castle}.'' This image is published
  under the cc-by-sa 2.0 license, independent of the license under which the
  work may be published in which this image and this contribution is embedded
  to. this licensing of this image does not touch the license of the remaining
  part of this contribution, as this contribution is considered to only make
  use of this image and is not derived from it.
\newblock
  \url{http://openstreetmap.org/?lat=49.01362&lon=8.40454&zoom=16&layers=B000F%
TF}.

\bibitem{Mawson1980}
A.~Mawson, ``{Is the concept of panic useful for scientific purposes?}'', in
  {\em International Seminar on Human Behavior in Fire Emergencies}, B.~Levin,
  ed., pp.~208--211.
\newblock NIST, Washington DC, 1980.

\bibitem{Mawson2005}
A.~Mawson, ``{Understanding mass panic and other collective responses to threat
  and disaster}'', {\em Psychiatry: Interpersonal and Biological Processes}
  {\bf 68} (2005) no.~2, 95--113.

\bibitem{_PED2008}
W.~Klingsch, C.~Rogsch, A.~Schadschneider, and M.~Schreckenberg, eds.,
  \href{http://dx.doi.org/10.1007/978-3-642-04504-2}{{\em {Pedestrian and
  Evacuation Dynamics 2008}}}.
\newblock Springer-Verlag, Berlin Heidelberg, 2010.
\newblock ISBN: 978-3-642-04504-2.

\end{thebibliography}\endgroup
%
%%%%%%%%%%%%%%%%%%%%%%%%%%%%%%%%%%%%%%%%%%%%%%%%%%%%%%%%%%%%%%%%%%%%%%%
\end{document}